\documentclass[conference]{IEEEtran}
\IEEEoverridecommandlockouts
\usepackage{eso-pic}
\usepackage{cite}
\usepackage{float}
\usepackage{dblfloatfix}
\usepackage{amsmath,amssymb,amsfonts}
\usepackage{algpseudocode}
\usepackage{algorithm}
\usepackage{graphicx}
\usepackage{textcomp}
\usepackage{xcolor}
\def\BibTeX{{\rm B\kern-.05em{\sc i\kern-.025em b}\kern-.08em
    T\kern-.1667em\lower.7ex\hbox{E}\kern-.125emX}}
\usepackage{subfig}
\usepackage{booktabs}
\usepackage{enumitem}
\usepackage[multiple]{footmisc}
\usepackage{balance}

\usepackage[latin9]{inputenc}
\usepackage{amsmath}
\usepackage{graphicx}

\begin{document}

\title{Chiron: Optimizing Fault Tolerance in QoS-aware Distributed Stream Processing Jobs}

\author{
\IEEEauthorblockN{Morgan K. Geldenhuys, Lauritz Thamsen,
and Odej Kao}
\IEEEauthorblockA{Technische Universit{\"a}t Berlin, Germany, \{firstname.lastname\}@tu-berlin.de}
}

\maketitle

\makeatletter

\AddToShipoutPicture*{\small \sffamily\raisebox{1.2cm}{\hspace{1.8cm}978-1-7281-6251-5/20/\$31.00 \copyright 2020 IEEE}}

\begin{abstract}
Fault tolerance is a property which needs deeper consideration when dealing with streaming jobs requiring high levels of availability and low-latency processing even in case of failures where Quality-of-Service constraints must be adhered to. Typically, systems achieve fault tolerance and the ability to recover automatically from partial failures by implementing Checkpoint and Rollback Recovery. However, this is an expensive operation which impacts negatively on the overall performance of the system and manually optimizing fault tolerance for specific jobs is a difficult and time consuming task.

In this paper we introduce \emph{Chiron}, an approach for automatically optimizing the frequency with which checkpoints are performed in streaming jobs. For any chosen job, parallel profiling runs are performed, each containing a variant of the configurations, with the resulting metrics used to model the impact of checkpoint-based fault tolerance on performance and availability. Understanding these relationships is key to minimizing performance objectives and meeting strict Quality-of-Service constraints. We implemented Chiron prototypically together with Apache Flink and demonstrate its usefulness experimentally.
\end{abstract}


\section{Introduction}
Distributed Stream Processing (DSP) systems are critical to the processing of vast amounts of data in real-time. It is here where events must traverse a graph of streaming operators to allow for the extraction of valuable information. There are many scenarios where this information is at its most valuable at the time of data arrival and therefore systems must deliver a predictable level of performance. Examples of such scenarios include IoT data processing, click stream analytics, network monitoring, financial fraud detection, spam filtering, news processing, etc. In order to keep up with the ever increasing data processing demands, DSP systems have the ability to scale horizontally across a cluster of commodity nodes to provide additional computation capabilities. However, as DSP systems scale to ever larger sizes, the probability of individual node failures increases and the need for more efficient fault tolerance mechanisms is attracting more attention.

The ability to continue operating in an environment where partial failures are to be expected is a core concern in distributed computing. Checkpoint and Rollback Recovery (CPR) is the most widely used fault tolerance strategy for DSP systems. It improves the overall reliability of streaming jobs executing within them as well as the integrity of results. This can be evidenced by its implementation in many of today's most popular streaming platforms such as Storm\cite{TTS+14}, Spark\cite{ZCF+10}, and Flink\cite{CKE+15}. This strategy involves creating a snapshot of the global state contained within the system and saving it. This is done so that, should a failure occur, the individual worker nodes can be instructed to stop, rollback to the latest save-point, and continue operating without the job failing\cite{KT87}.

However, the CPR process is resource intensive and invariably introduces performance overhead that only grows more significant as requirements increase. This is mainly due to three factors: the replication, transport, and storage of state that needs to be saved at regular intervals; \textit{event-time} processing; and maintaining any fault tolerance guarantees, i.e. \textit{end-to-end exactly once} semantics, while interacting with external systems
. For DSP systems which use CPR, selecting an optimal checkpoint interval is key to ensuring high efficiency in streaming applications. Ultimately, there is a trade-off between the performance overhead of regularly saving the global state of the system and the runtime costs of failure recovery \cite{GTG+19}. In essence, setting the checkpoint frequency too low risks longer recovery times and therefore allows for weaker availability, while setting the checkpoint frequency too high impacts on the overall performance of normal stream processing.

In this paper we address the problem faced by DSP pipeline operators whose jobs must adhere to strict Quality of Service (QoS) constraints where the scaling out of resources is undesired or infeasible. It answers the question: How can a DSP be configured so that it provides an optimal end-to-end latency, while ensuring recovery from failures within a given time limit? We propose \emph{Chiron}, an automatic profiling and runtime prediction approach which models these relationships and then employ optimization techniques to find good fault tolerance configuration settings for DSP jobs with strict QoS requirements. To the best of our knowledge, our approach is novel in its attempt to model both the performance and availability of DSP jobs through profiling and recommend an optimal checkpoint interval based on user-defined QoS constraints for improving overall performance. We also contribute to the body of knowledge by presenting an in-depth analysis of the most common failure scenario, a new heuristic for calculating the total time a DSP job will be unavailable, and the results of two experiments conducted in support of our hypothesis.


The remainder of the paper is structured as follows: Section II presents an analysis of the problem; Section III introduces a new heuristic for measuring availability; Section IV presents our approach to profiling, modeling, and optimization; Section V describes our evaluation through experiments; Section VI the related work; and Section VII summarizes our findings.

\section{Problem Analysis}

In this section we examine what happens when a failure occurs and how a DSP system would respond. This allows for a greater understanding of the CPR process and any considerations which need to be taken into account while attempting to find good configurations. The common case of a worker node failing during execution of a DSP job is used.

\begin{figure}[ht]
    \centering
    \includegraphics[width=0.98\linewidth]{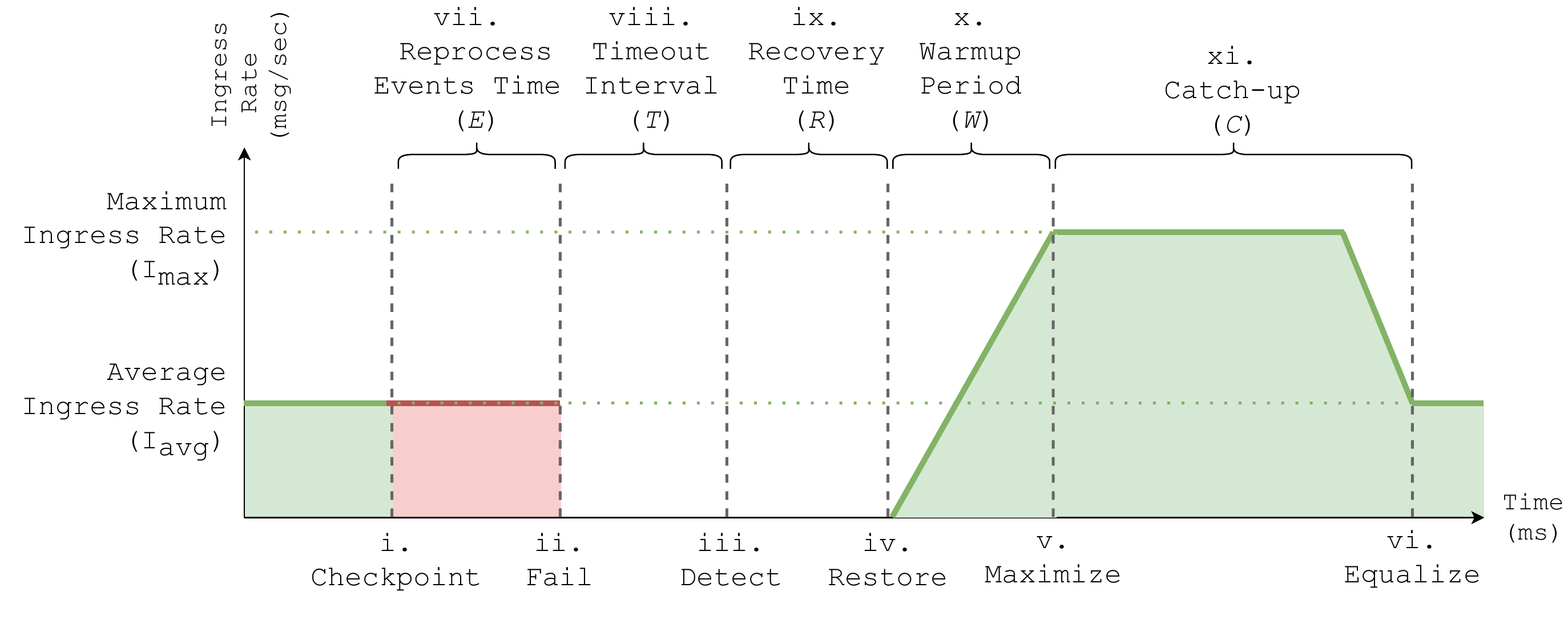}
    \caption{Worker node failure scenario.}
    \label{nodefailure}
\end{figure}

A master-worker architecture is assumed for this scenario, processing occurs at event time, and events are consumed from an external source providing exactly-once fault tolerance guarantees. Figure \ref{nodefailure} illustrates the scenario of a worker node failing silently, i.e. crashing without sending a notification. It is therefore the responsibility of the system to determine that one of the workers has stopped prematurely, revert the system back into a consistent state, and catch-up to the latest event stream offset. We describe events as they would naturally occur, chronologically from left to right (points \textit{i}. to \textit{v}.) along the $x$-axis. The $y$-axis represents the ingress rate, i.e. the cumulative frequency at which messages enter the source operators of the dataflow. The points in time are:

\begin{enumerate}[leftmargin=*,topsep=0.5em]
    \itemsep0.5em

    \item [i.] \verb|Checkpoint|: The checkpoint process has completed successfully and the distributed snapshot of the global state is saved to disk along with the current event stream offset of the external source. No messages processed prior to this point will need to be reprocessed in the event of a failure.
    
    \item [ii.] \verb|Fail|: At some point before the next checkpoint completes successfully, a worker node fails silently. The system has not yet detected this failure and will attempt to process and perform checkpoints as per normal. Checkpoints will ultimately fail as the consensus between cluster nodes cannot be reached. Events processed before this point but after the last checkpoint will need to be reprocessed.
    
    \item [iii.] \verb|Detect|: At this point, the system has detected the failure. Generally DSP systems use a heartbeat protocol to timeout non-responsive nodes. Once the failure has been detected, all running tasks are cancelled and the worker nodes are instructed to rollback to the last checkpoint.
    
    \item [iv.] \verb|Restore|: At this point, all worker nodes have reverted to their previously saved state and processing will begin from the last committed event stream offset. As this offset is further back in time than the current timestamp, the system will attempt to "catch up" and the ingestion rate will rapidly increase up until the maximum processing capacity of the system has been reached.
    
    \item [v.] \verb|Maximize|: At this point, the maximum processing capacity of the system has been reached, events will be processed at this fairly stable rate until the job has caught-up. Ingress rates are influenced by the number of available resources. 
    
    \item [vi.] \verb|Equalize|: At this point, the DSP job has processed the backlog of accumulated events and the job reaches equilibrium with the average ingress rate. Processing continues as per normal and the extra resources which are no longer necessary are released.

\end{enumerate}



Fault tolerance is an inherently expensive operation which requires resources in order for a DSP job to be able to recover. After checkpointing, a distributed snapshot of the global state is contained at the worker nodes which then needs to be copied over the network to be stored in case it is required later. This is a network intensive operation and the frequency with which this process is initiated plus the size of each snapshot will have a negative impact on the overall end-to-end latencies. This is especially important to consider in instances where performance requirements are defined in SLAs. When a failure occurs, the DSP system will be unavailable for a period of time. During this downtime, events will accumulate at the external source(s). This includes those which will need to be reprocessed as a result of the rollback. Importantly, the point at which processing resumes should not be considered the point at which the system is once again available, the backlog of accumulated events will first need to be dealt with. We define \textit{Total Recovery Time} ($TRT$) as the time required for a DSP job to catch-up to the latest offset of the incoming event stream from the point at which the failure occurred. It is the metric by which we measure availability in DSP jobs and is reliant on a number of factors. Referring to Figure \ref{nodefailure}, the $TRT$ is composed of the following time periods:

\begin{enumerate}[leftmargin=*,topsep=0.5em]
    \itemsep0.5em
    
    \item [vii.] \verb|Reprocess Events Time| (\textit{E}): Represents the time needed to reprocess uncommitted events arriving after the last successful checkpoint completed but before the failure was detected. It is not possible to predict exactly how many events are to be reprocessed nor how long this would take as it is dependent on knowing the exact timestamp of when the failure would occur after the last successful checkpoint. However, a minimum, median, and maximum number of messages can be estimated based on the time between checkpoints and the average rate of messages being processed per second. 
    
    \item [viii.] \verb|Timeout Interval| (\textit{T}): Represents the time the DSP system will wait before declaring a non-responsive worker node as failed. In most systems this is based on the \textit{heartbeat timeout} configuration setting.
    
    \item [ix.] \verb|Recovery Time| (\textit{R}): Represents the length of time the DSP system will take to go from an inconsistent state back into a consistent state after a failure has been detected. Recovery time is influenced by a number of factors, both specific to the DSP job (i.e. snapshots size) as well as the features of the DSP (i.e. differential checkpoints, maintaining local copies of worker state, restart strategies, etc.).
    
    \item [x.] \verb|Warm-up Period| (\textit{W}): Represents the time it takes for the ingress rate across all source operators to collectively increase from zero to the maximum. This is reliant on a number of factors, primarily of which are the read rates of external sources and the maximum capacity of system. 
    
    \item [xi.] \verb|Catch-up| (\textit{C}): Represents the time the DSP job will take to process the backlog of messages accumulated while the system was unavailable. The DSP system will use the maximum processing capacity available in an attempt to catch-up and the more physical resources that are available, the faster this process will take to complete. The catch-up time period is directly related to the average ingress rate and the collective time spent in the proceeding phases.
    
\end{enumerate}

Calculating an accurate $TRT$ is not a trivial matter as it is highly dependent on the operational characteristics of the specific DSP job being executed and a number of factors which are not known prior to execution time. However, for jobs where there are defined QoS targets and the system should be available and caught-up after a finite amount of time, an estimate is needed for configuring systems well.

\section{Estimating the Total Recovery Time}

Building upon the analysis which was presented in the previous section, we introduce a new heuristic for estimating the minimum, average, and maximum $TRT$ per data point gathered through profiling runs. Referring to Figure \ref{nodefailure}, the $TRT$ is modeled as a decreasing geometric sequence where the $1^{st}$ term is composed of the time periods $vii$ through $x$ multiplied by a common ratio $U$. The first time period $E$ is directly related to the \textit{checkpoint interval} ($CI$) configuration. The $CI$ is defined as the frequency with which the checkpoint process is initiated and is measured in milliseconds. Since we cannot predict exactly when the failure will occur between checkpoints, we can take a best, average, and worst case estimate. Therefore, zero in the best case scenario, the length of the $CI$ divided by two in the average case, and the full length of the $CI$ in the worst case. The second time period $T$ is based on the \textit{heartbeat timeout} configuration variable which determines how long the system will wait until a node is timed out. In order to calculate the common ratio, a measure of the processing capacity utilization is required. We know that the system will use the maximum processing capacity of the system on top of what is already being utilized to catch-up after a failure has occurred. Therefore, knowing the average rate at which messages are processed ($I_{avg}$) and the maximum rate at which the system is capable of processing ($I_{max}$) can be combined to formulate the processing capacity utilization as a percentage (\textit{U}) as seen in Equation \ref{eq:1}.

\begin{equation}
  \label{eq:1}
  U = \frac{I\textsubscript{avg}}{I\textsubscript{max}}
\end{equation}





Knowing the $1^{st}$ term and the common ratio allows us to approximate how long it takes to process events arriving while the system was inoperable. However, this does not account for the events which arrived while the system was performing this catch-up, nor the time it takes to catch-up on the catch-up, or to catch-up on the catch-up of the catch-up, etc. It is for this reason that we model our heuristic as a decreasing geometric series. The formula for this series can be seen in Equation \ref{eq:2}. 

\begin{equation}
  \label{eq:2}
  C(n) =
  \begin{cases}
    (E+T+R+W)\cdot{U} & \text{if $n=1$} \\
    C(n-1)\cdot{U} & \text{if $n>1$}
  \end{cases}
\end{equation}

\begin{figure*}[!h]
    \centering
    \includegraphics[width=1.0\textwidth]{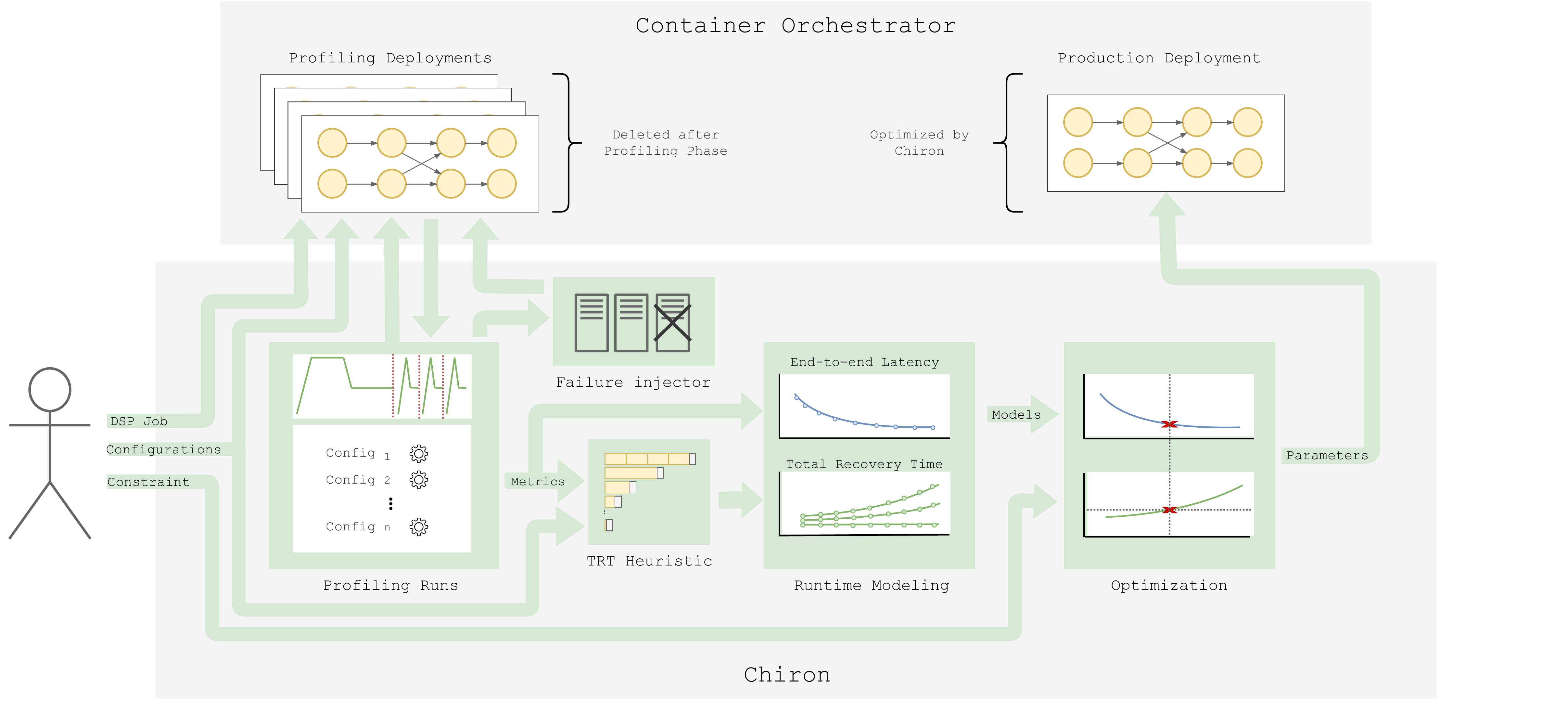}
    \caption{Overview of our approach with Chiron and its interactions with users and systems.}
    \label{overview}
\end{figure*}

In order to determine the $TRT$, the sum of this geometric sequence needs to be calculated. Doing so requires that the $n^{th}$ term be known. It is important to note that the number of terms in a decreasing geometric sequence will approach infinity as outputs tend to zero and therefore a stopping condition must be established. For our calculations, we recommend choosing the first $n$ resulting in a value less than one. The formula for finding the $n^{th}$ term thus can be calculated using Equation \ref{eq:3} executed in an iterative loop $1..n$.

\begin{equation}
  \label{eq:3}
  a\textsubscript{n} = (E+T+R+W)\cdot{U^{(n-1)}}
\end{equation}

With the number of terms $n$, the $1^{st}$ term of $C(n)$, and the common ratio $U$; a formula for finding the sum of a geometric sequence can be derived. This formula can be seen in Equation \ref{eq:4}. The output will equate to the time taken to catch-up to the latest offset of the event stream from the point when the system recovered its ability to start processing events once more.

\begin{equation}
\label{eq:4}
  S\textsubscript{n} = \frac{(E+T+R+W)(1-U^{n})}{(1-U)}
\end{equation}

With the $S_{n}$ being known, the $TRT$ can finally be calculated by combining it with the time periods of when the system was in an inconsistent state after the failure, i.e. $T$ and $R$. The formula for this can be seen in Equation \ref{eq:5}.

\begin{equation}
\label{eq:5}
  TRT = T + R + S(n)
\end{equation}

With the ability to estimate the $TRT$, we are now able to proceed to modeling the behaviors of the DSP job as a function of the $CI$.

\section{Approach}

In DSP systems, configuration has a direct impact both on performance and availability. Yet, determining exactly how much of an impact is hard to ascertain. Regarding CPR fault tolerance, the most important configuration setting to take into consideration is the \textit{checkpoint interval} ($CI$). Our approach, Chiron, automatically selects an optimized $CI$, given a user defined QoS constraint. This is achieved by following three consecutive steps: 1. \textit{Profiling}, where we make use of a number of enabling technologies to efficiently gather metrics for DSP jobs executing with different $CI$ configuration settings; 2. \textit{Modeling}, where we take these metrics and the estimation for the $TRT$ in order to model relationships for both performance and availability in terms of the $CI$, and; 3. \textit{Optimization}, where we take these models and based on the user requirements, optimize for the chosen objective. Chiron is intended to be executed either for a short time at the start of a DSP job or periodically to ensure current runtime conditions are taken into account. A streaming job's configuration could, for instance, be validated using Chiron once every six hours and testing a set of configurations using the current input stream for ten minutes. An overview of Chiron can be seen in Figure \ref{overview}.

\subsection{Profiling}


To meet the performance modeling requirements of DSP jobs, we use a technique for gathering metrics in environments that closely mirrors realistic conditions. By employing two key enabling technologies, i.e. OS-virtualization and container orchestration; we are able to replicate multiple pipelines in parallel while ensuring isolation from one another. At the same time, identical copies of the DSP job can be deployed within their own separate self-contained environments while possessing a unique variation of the system configurations. These technologies, when combined with Infrastructure-as-Code processes, provide a mechanism for quickly reproducing any environment. Our previous work, Timon\cite{GTG+19}, introduced a tool for automatically achieving these goals. To ensure results are comparable and as close to how they would be under realistic conditions, each parallel deployment consumes the same data stream during profiling under normal expected loads. In order to select a good set of input configuration variables for each parallel deployment, the solution space is evenly explored by selecting a minimum and maximum value for the $CI$ after which a set of equidistant values are calculated between these extremes. The following metrics are gathered from each of the parallel deployments:

\begin{itemize}[leftmargin=*]
    \setlength{\itemsep}{3pt}
    
    \item{\verb|Average Ingress Rate| ($I_{avg}$)}: Measured in events per second, this value represents the average rate at which events enter the source operators under normal load.
    
    \item{\verb|Maximum Ingress Rate| ($I_{max}$)}: Measured in events per second, this value represents the maximum rate at which events can be processed. This value can be estimated by performing load testing.
    
    \item{\verb|Average End-to-End Latency| ($L_{avg}$)}: Measured in milliseconds, this value represents the average time taken for an event to traverse the execution graph from the source to the sink operator. Windowing periods are not considered as part of this measurement. Average system performance is determined by this value and is directly affected by the fault tolerance mechanism, i.e. the larger in size and frequency of snapshots, the greater the impact on system resources.
    

    \item{\verb|Average Recovery Time| ($R_{avg}$)}: Measured in milliseconds, this value represents the average recovery time as described in the Problem Analysis section. It can be measured by sequentially injecting failures during profiling and averaging the time taken to recover.
    
    \item{\verb|Average Warm-up Time| ($W_{avg}$)}: Measured in milliseconds, this value represents the average time taken for the ingress rate to increase from zero to the maximum processing capacity. Like with $I_{max}$, this value can be estimated by performing load testing.

\end{itemize}

\subsection{Modelling}

\begin{figure}[!h]
\centering
\subfloat[\textit{CI} vs. Avg. End-to-end Latency.]{
  \includegraphics[width=40mm]{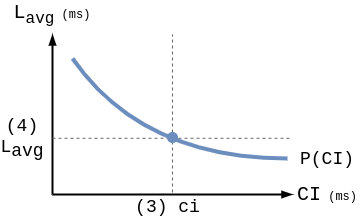}
}
\subfloat[\textit{CI} vs. Total Recovery Time.]{
  \includegraphics[width=40mm]{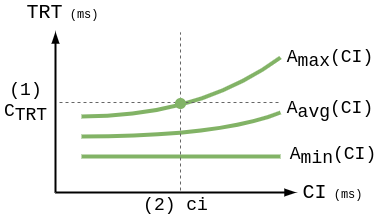}
}
\caption{Modeling and optimization with availability constraint.}
\label{modelingoptimization}
\end{figure}

In our approach, we propose the generation of two two-population models. The first for modeling performance as a predictor of $L_{avg}$ and the second for modeling availability as a predictor of $TRT$. Concerning performance, data points are taken directly from profiling runs and used for modeling. The function graph $P(CI)$ in Figure \ref{modelingoptimization}(a) illustrates our estimate for the relationship between the independent variable $CI$ and the dependent variable $L_{avg}$. Predictably, the gradient of the curve flattens as $CI$ increases due to its impact on performance lessening substantially as checkpoint frequency decreases. For availability, the $TRT$ estimate generated by the heuristic can be used to model a family of functions. The function graphs $A_{min}(CI)$, $A_{avg}(CI)$, and $A_{max}(CI)$ in Figure \ref{modelingoptimization}(b) illustrate these relationships. Here we can see that for each $CI$ inputted, three corresponding outputs can be returned for the best, average, and worst cases. Logically, lower $CI$ frequencies will have a more profound impact on the $TRT$ than higher ones. With these functions we can proceed to the final optimization.


\subsection{Optimization}

The goal of our approach is to predict a $CI$ which will configure the fault tolerance mechanism in such a way as to optimize for performance while setting an upper bound on the time the system will be unavailable in the event of a failure. As such, the user will be required to provide a QoS constraint $C_{TRT}$ defining the upper bound on the $TRT$. With this input and the outputs of the modeling process, all necessary components are at hand to perform the optimization step. The choice of whether to plan for the worst or the average case is left up to the user. The first step in the process is to find the corresponding $CI$ value of the associated $C_{TRT}$ constraint. Using the inverse of the selected $A(CI)$ function the constraint can be inputted to find the $CI$ value used for system configuration. The next step is to find the predicted minimum objective using the alternate function and finally return all three values, i.e. $CI$, $C_{TRT}$, and $L_{avg}$. 



\section{Evaluation}

Now we show that using Chiron is both practical and beneficial for distributed stream processing through experiments.


\subsection{Experimental Setup}

Our experimental setup consisted of a 10-node Apache Kafka cluster \cite{KNR11} and a 50-node Kubernetes \cite{VPK+15} and HDFS cluster \cite{VMD+13}. Node specifications and software versions are summarized in Table \ref{clusterspecs}. A single switch connected all nodes. For both experiments, 11 Flink clusters were instantiated to perform parallel profiling runs across 11 checkpoint interval settings ranging from a minimum of 1000ms to a maximum of 60,000ms. Parallelism for all jobs was set to 24. Each Flink cluster consisted of 1 job master in high-availability mode and 27 workers. All Flink workers were created with 1 task slot and 4GB of memory. A total of 5 profiling runs were conducted for each experiment with the median resulting values being selected for modeling. Prometheus\footnote{ https://prometheus.io, Accessed: Aug 2020} was used for the gathering of metrics. A user-defined $TRT$ constraint ($C_{TRT}$) was configured for each experiment. This constraint is defined as the maximum time the job should require before being available again in the event of a failure.

\begin{table}[ht]
\centering
    \begin{tabular}[t]{rp{0.65\linewidth}}
        \toprule
        Resource&Details\\
        \midrule
        OS&Ubuntu 18.04.3\\
        CPU&Quadcore Intel Xeon CPU E3-1230 V2 3.30GHz\\
        Memory&16 GB RAM\\
        Storage&3TB RAID0 (3x1TB disks, linux software RAID)\\
        Network&1 GBit Ethernet NIC\\
        Software&Java v1.8, Flink v1.10, Kafka v2.5, ZooKeeper v3.5, Docker v19.03, Kubernetes v1.18, HDFS V2.8, Redis v5.08, Prometheus v2.17, Pumba v0.7 \\
        \bottomrule
    \end{tabular}
\caption{Cluster Specifications}
\label{clusterspecs}
\end{table}%

To determine the maximum processing capacity of each job, profiling runs were initiated from an earlier timestamp resulting in approximately 10 minutes of catch-up time. Kafka was configured with 24 partitions per topic and was not a limiting factor in the maximum input throughput at any point. Regarding latencies, averages were taken over the 0.999 percentile in order to filter outliers during normal failure free operations. Three failures were injected per job execution via the Pumba chaos testing tool and the average time taken for the job to restart was calculated. Actual $TRT$s were also measured independently in order to validate the outputs of the modeling process. Lastly, second order ($k=2$) polynomial linear regression was selected for curve fitting across all models.
 
\begin{figure*}
\centering
\subfloat[IoTDV Experiment: $P(CI)$.]{
  \includegraphics[width=42mm]{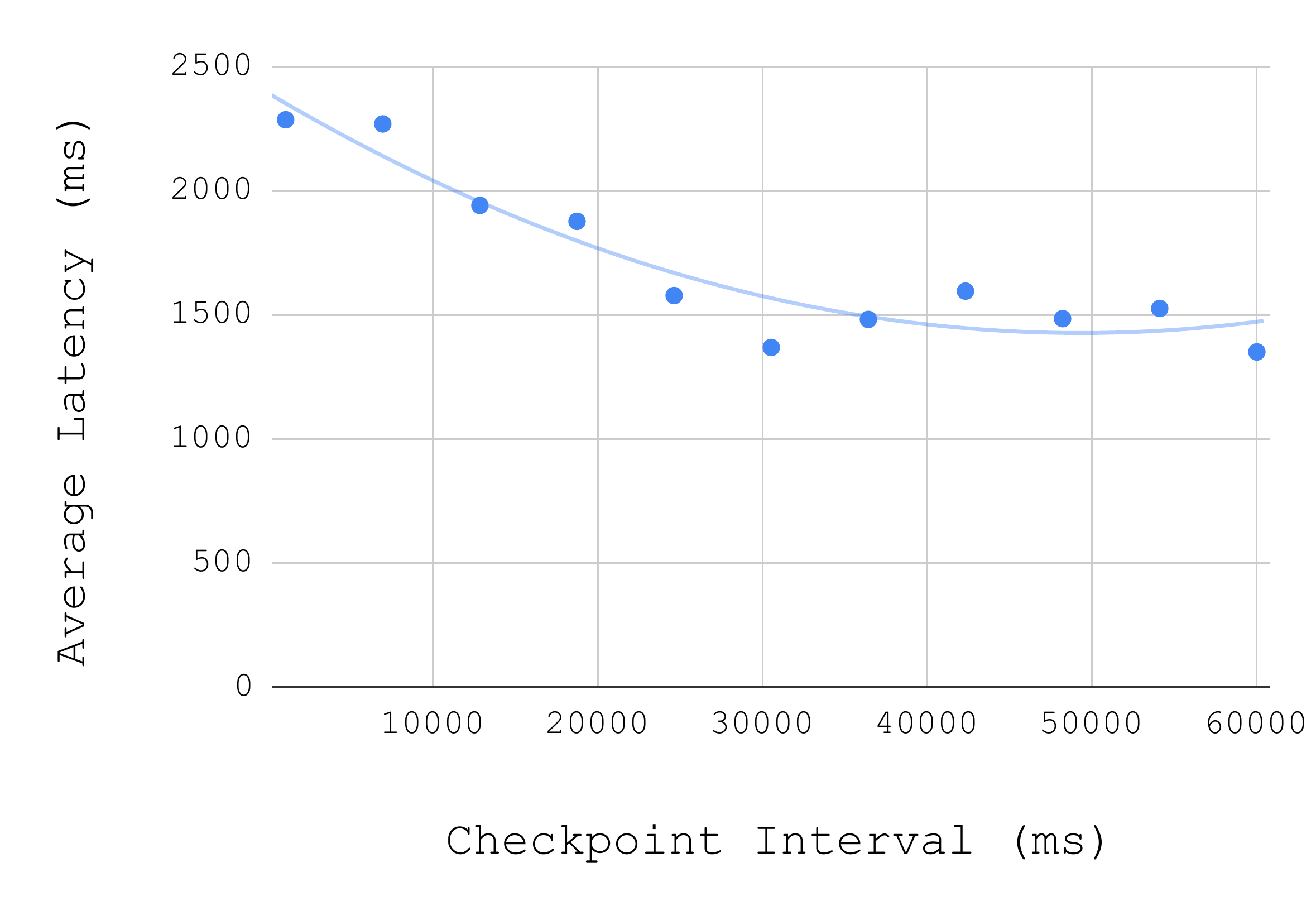}
}
\subfloat[IoTDV Experiment: $A_{max}(Ci)$, $A_{avg}(CI)$, and $A_{min}(CI)$.]{
  \includegraphics[width=42mm]{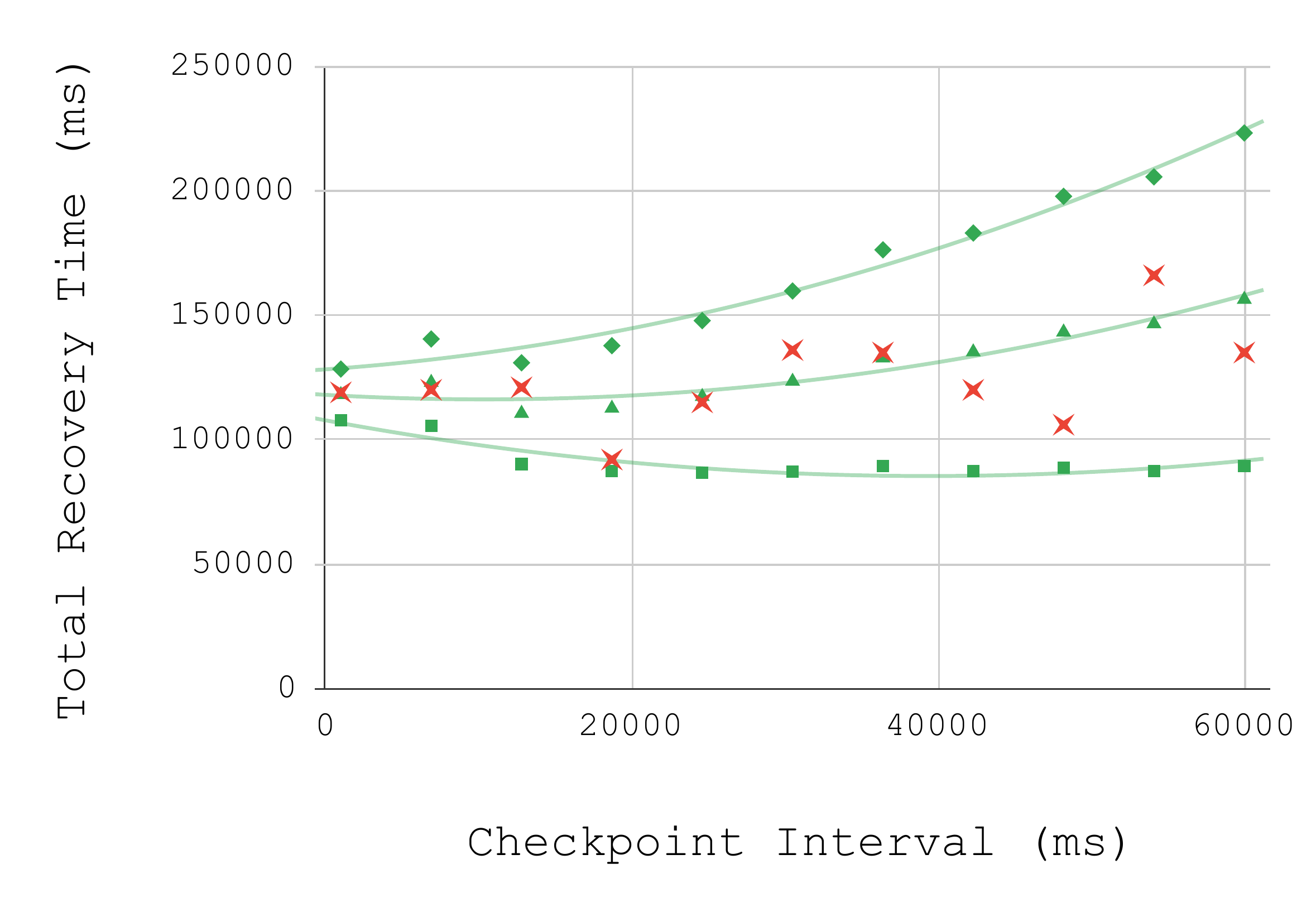}
}
\subfloat[YSB Experiment: $P(CI)$.]{
  \includegraphics[width=42mm]{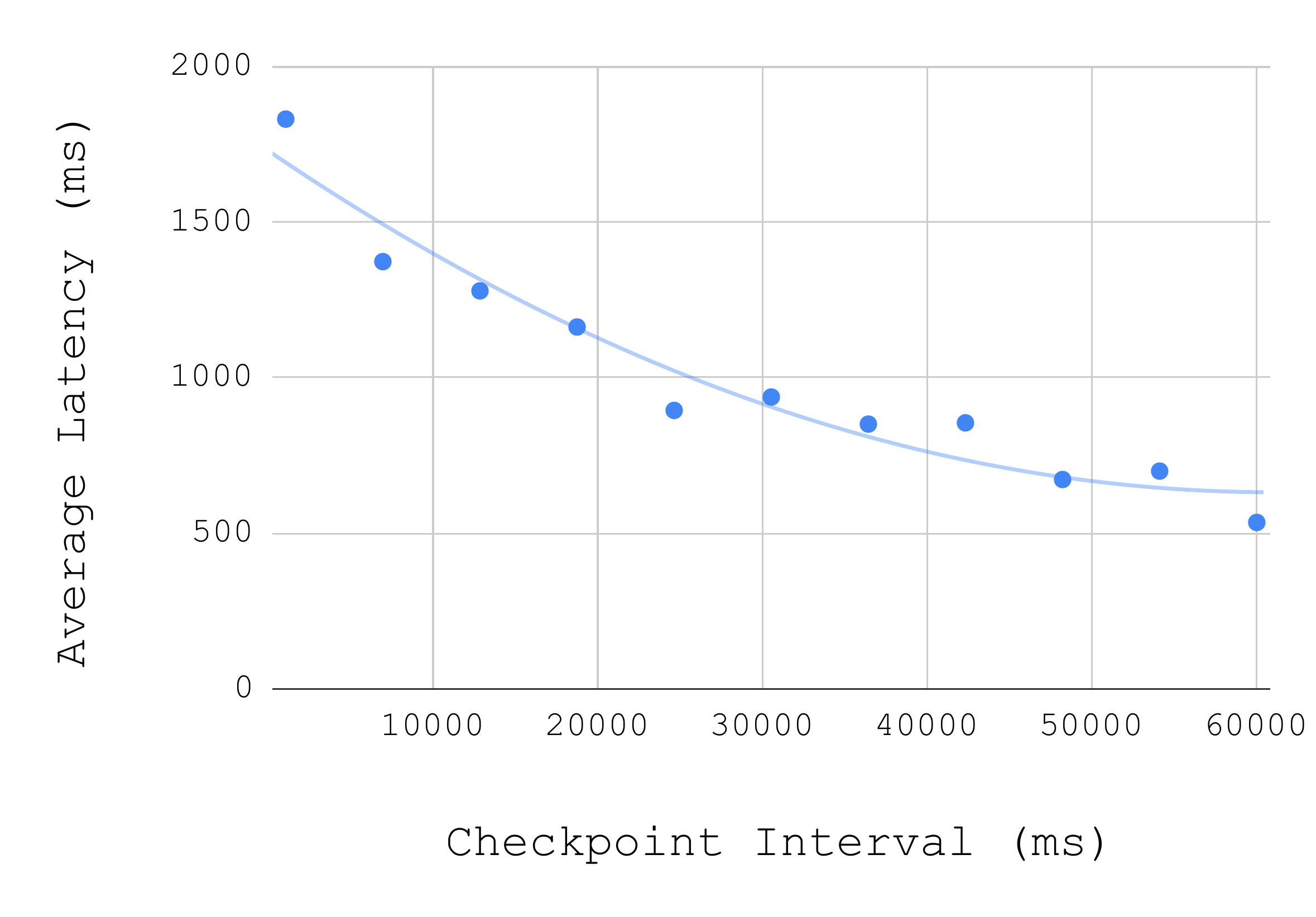}
}
\subfloat[YSB Experiment: $A_{max}(CI)$, $A_{avg}(CI)$, and $A_{min}(CI)$.]{
  \includegraphics[width=42mm]{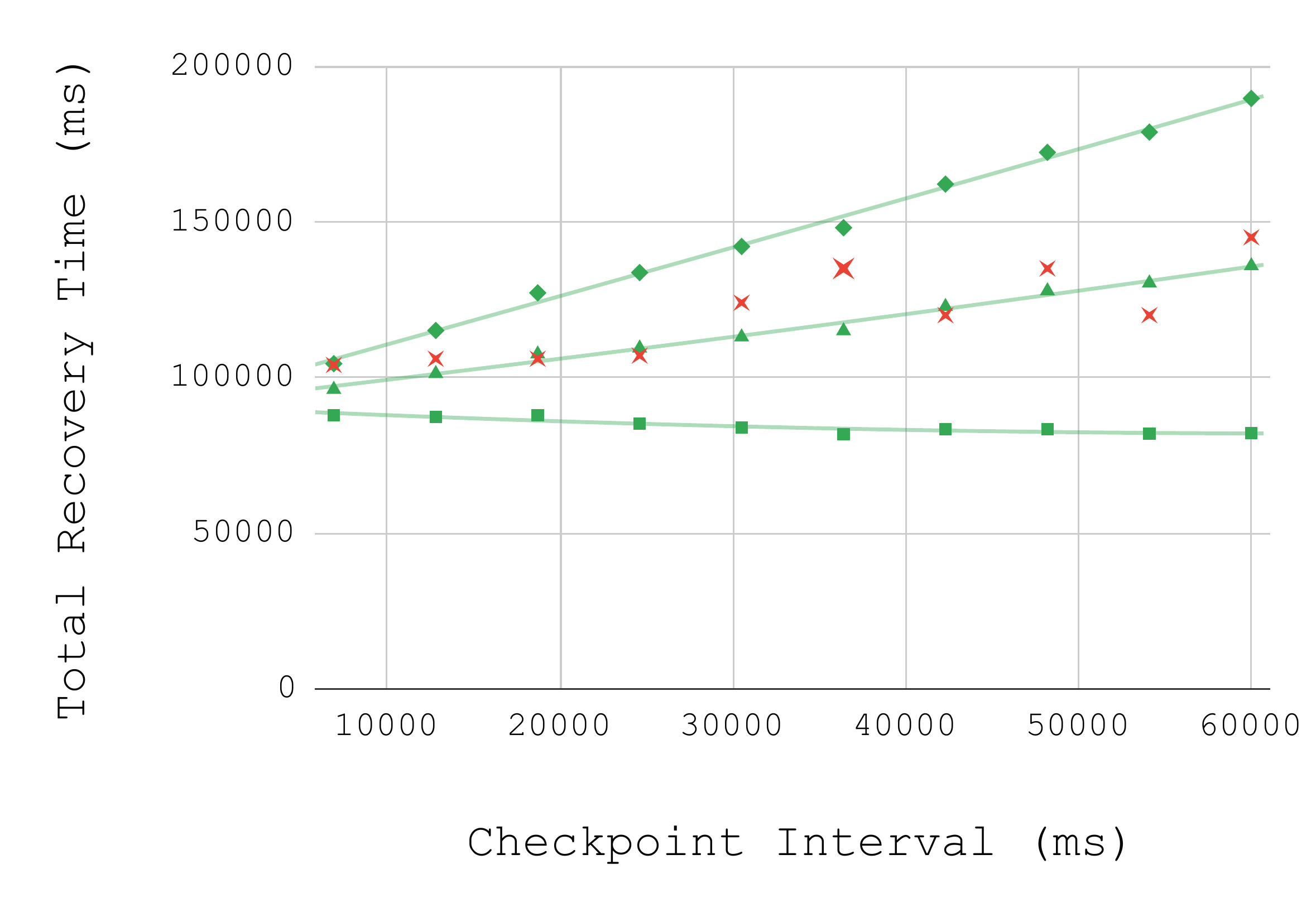}
}
\caption{Experimental outputs of modeling process.}
\label{modelingoutputs}
\end{figure*}
 
\subsection{Streaming Jobs} 
 
Two experiments were conducted to evaluate the usefulness of Chiron. Source code for these experiments can be found on github\footnote{https://github.com/morgel/IoTDV-experiment}\footnote{https://github.com/morgel/YSB-experiment}. The first was the IoT Delivery Vehicles (IoTDV) Experiment where a simulator generates 500,000 delivery vehicle events per second which are submitted to Kafka to await processing. Each event contains information about the vehicle's geo-location, speed, and crucially whether or not the vehicle is self-driven. The job consisted of the following streaming operations: read an event from Kafka; deserialize the JSON string; filter update events not within a certain radius of a designated observation geo-point where delivery vehicles are of a particular type, i.e. self-driving; take a 10 second window where all update messages are of the same vehicle ID and calculate the vehicles average speed; generate an alarm for vehicles which have exceeded the speed limit; enrich notification with vehicle type information from data stored in system memory and write it back out to Kafka. The second experiment is based on the Yahoo Streaming Benchmark (YSB)\footnote{ https://yahooeng.tumblr.com/post/135321837876/benchmarking-streaming-computation-engines-at, Accessed: Aug 2020}. It implements a simple streaming advertisement job where there are multiple advertising campaigns and multiple advertisements for each campaign. The job consists of the following operations: read and event from kafka; deserialize the JSON string; filter out irrelevant events (based on type field), take a projections of the relevant fields (ad\_id and event\_time),  join each event by ad\_id with its associated campaign\_id stored in Redis; take a 10 second windowed count of events per campaign and store each window in Redis along with a timestamp of when the window was last updated. For the purposes of our experiments, we modified the Flink benchmark by enabling checkpointing and replacing the handwritten windowing functionality with the default Flink implementation. Although doing so does decrease update frequency to the length of each window, results should be accurate and more interesting for our experiments due to the accumulated windowing operator state at each node.

\begin{table}
\caption{IoTDV Experiment Results.}
\centering
    \subfloat[Coefficient of Determination.] {
    \begin{tabular}{p{21mm}|p{9mm}p{9mm}p{9mm}p{9mm}} \toprule
        {} & {$P$} & {$A_{max}$} & {$A_{avg}$} & {$A_{min}$} \\ \midrule
        {R-Squared} & 0.891 & 0.98 & 0.934 & 0.819 \\ \bottomrule
    \end{tabular}
    }
    \newline
    \subfloat[Experimental Outputs of Optimization Process] {
    \begin{tabular}{p{21mm}|p{9mm}p{16mm}p{16mm}} \toprule
        {} & {} & {$CI$} & {$L_{avg}$} \\ \midrule
        {Values (ms)} & {} & 41,581 & 1447 \\ \bottomrule
    \end{tabular}
    }
    \newline
    \subfloat[Error Analysis of Experimental Outputs against Actual Values.] {
    \begin{tabular}{p{21mm}|p{5mm}p{5mm}p{5mm}p{5mm}p{5mm}|p{5mm}} \toprule
        {Observation} & {\#1} & {\#2} & {\#3} & {\#4} & {\#5} & {Target} \\ \midrule
        {Actual $TRT$ (s)} & 120 & 114 & 105 & 105 & 151 & 180 \\
        {$C_{TRT} > TRT$} & true & true & true & true & true \\ \midrule
        {Actual $L_{avg}$ (ms)} & 1625 & 1537 & 1501 & 1474 & 1615 & 1447 \\
        {Percent Error (\%)} & 12.30 & 6.22 & 3.73 & 1.87 & 11.61 \\ \bottomrule
    \end{tabular}
    }
    \label{iotresults}
\end{table}
 
\begin{table}
\caption{YSB Experiment Results.}
\centering
    \subfloat[Coefficient of Determination.] {
    \begin{tabular}{p{21mm}|p{9mm}p{9mm}p{9mm}p{9mm}} \toprule
        {} & {$P$} & {$A_{max}$} & {$A_{avg}$} & {$A_{min}$} \\ \midrule
        {R-Squared} & 0.942 & 0.996 & 0.989 & 0.861 \\ \bottomrule
    \end{tabular}
    }
    \newline
    \subfloat[Experimental Outputs of Optimization Process.] {
    \begin{tabular}{p{21mm}|p{9mm}p{16mm}p{16mm}} \toprule
        {} & {} & {$CI$} & {$L_{avg}$}\\ \midrule
        {Values (ms)} & {} & 35,195 & 826\\ \bottomrule
    \end{tabular}
    }
    \newline
    \subfloat[Error Analysis of Experimental Outputs against Actual Values.] {
    \begin{tabular}{p{21mm}|p{5mm}p{5mm}p{5mm}p{5mm}p{5mm}|p{5mm}} \toprule
        {Observation} & {\#1} & {\#2} & {\#3} & {\#4} & {\#5} & {Target}\\ \midrule
        {Actual $TRT$ (s)} & 106 & 107 & 105 & 130 & 105 & 150\\
        {$C_{TRT} > TRT$} & true & true & true & true & true \\ \midrule
        {Actual $L_{avg}$ (ms)} & 906 & 917 & 866 & 871 & 957 & 826 \\
        {Percent Error (\%)} & 9.69 & 9.92 & 4.62 & 5.17 & 13.69 \\ \bottomrule
    \end{tabular}
    }
    \label{yahooresults}
\end{table}

\subsection{Experimental Results}

A $C_{TRT}$ of 180s and 150s was selected for the IoTDV Experiment and YSB Experiment, respectively. Based on the metrics gathered during profiling runs, 4 functions were generated as part of the modeling process for each experiment. These included: $P(CI)$ for performance modeling as illustrated in Figures \ref{modelingoutputs}(a) and \ref{modelingoutputs}(c); and the family of functions $A_{max}(CI)$, $A_{avg}(CI)$, and $A_{min}(CI)$ for availability modeling as illustrated in Figures \ref{modelingoutputs}(b) and \ref{modelingoutputs}(d). $R^2$ values for these functions can be seen in Tables \ref{iotresults}(a) and \ref{yahooresults}(a). For these experiments, $A_{max}(CI)$ was used for availability calculations. As the goal of these experiments was to predict a $L_{avg}$ and $CI$ based on the user-defined $C_{TRT}$ constraint, Tables \ref{iotresults}(b) and \ref{yahooresults}(b) detail the outputs of the optimization process. Here we can see for the IoTDV Experiment a $CI$ of 41s was predicted with an $L_{avg}$ of 1447ms. Regarding the YSB Experiment, a $CI$ of 35s was predicted with an $L_{avg}$ of 826ms. In order to evaluate the accuracy of our approach, we performed an error analysis of the experimental values against actual observations. As part of the evaluation process, each experiment was executed 5 times with the predicted $CI$ configuration settings and metrics were recorded. Tables \ref{iotresults}(b) and \ref{yahooresults}(b) tabulate the results of the error analysis performed for both experiments. Here we can see that all experimental values fall within 15\% of the actual observed values for the $L_{avg}$ and all measured $TRT$ values were less than the $C_{TRT}$. This indicates that this approach is able to closely predict actual runtime conditions. Additionally, in order to verify the accuracy of the $TRT$ heuristic, during profiling runs the total time between when the failure was injected and the point when all accumulated lag on the source operators had been caught-up was measured independently during profiling runs. The red X-marks on Figures \ref{modelingoutputs}(b) and \ref{modelingoutputs}(d) represent the median values for these observations. For both graphs, the majority of observations were within the range between the minimum and maximum estimates and, as median values, find themselves plotted in close proximity to $A_{avg}(CI)$. This is a good indication that the heuristic is able to predict that further observations will fall between these ranges.

\section{Related Work}

Work most related to our own includes research aimed at adaptive checkpointing. Multi-level checkpointing has been proposed to resolve the issue of checkpoint/recovery overhead \cite{GKA+10,GMC+10,LPW+14,MBM+10,BTK+11,KMI+12}. This allows the use of different checkpointing levels, making it more flexible than traditional single-level approaches which are usually limited to one storage type. It considers multiple failure types with each having a different checkpoint and restart cost associated. For instance, \cite{DRV+17} proposes a two-level checkpointing model whereby checkpoint level 1 deals with errors with low checkpoint/recovery overheads such as transient memory errors, while checkpoint level 2 deals with hardware crashes such as node failures. However, these approaches are specific to HPC environments and need to be adapted before use in DSP systems. Other approaches have been proposed which optimize the configuration parameters of this mechanism by finding an optimal $CI$ to improve performance. Some focus on determining the mean time to failure (MTTF) of cluster nodes and then adaptively fitting a $CI$ which minimizes the time lost due to failure \cite{Y74,D03,D06}. However, these approaches rely on jobs having a finite execution time as part of their calculations which is more appropriate to HPC clusters and batch processing workloads than stream processing jobs where completion times are unbounded. A more recent approach specific to DSP systems likewise incorporates failure rates in an attempt to fit a $CI$ based on the MTTF \cite{Jayasekara2020AUM}. Our work differs in that Chiron follows a profiling approach intended to be executed either for a short time at the start of a DSP job or periodically as apposed to continuous execution for runtime optimizations. Additionally, we focus on performance and availability modeling rather than MTTF.

\section{Conclusion}

Chiron models the performance and availability behaviors of DSP jobs based on metrics gathered from profiling runs. It does so in order to predict an optimal $CI$ configuration setting which, based on a user-defined maximum total recovery time QoS constraint, will minimize the performance impact of enabling fault recovery. Profiling runs are conducted in parallel with the aid of two key enabling technologies, i.e. OS-virtualization and container orchestration. Chiron introduces a new heuristic for predicting the minimum, average, and maximum \textit{total recovery time}---a measure of the total time from the point when a failure occurs to the point when the job has once again caught-up to the head of the incoming event stream. Our approach is aimed at use cases where the data stream is processed at event time and characterized by a fairly stable average input throughput. Although some variance in this rate is to be expected, use cases where the number of events entering the source operators is essentially random or fluctuates wildly over time is not conducive to finding optimal configurations through profiling runs. Future work may focus on using the metrics gathered during profiling to optimize the fault tolerance mechanism of running analytics pipelines.


\section*{Acknowledgment}

This work has been supported through grants by the German Ministry for Education and Research (BMBF) as BIFOLD (funding mark 01IS18025A) and WaterGridSense 4.0 (funding mark 02WIK1475D).

\balance

\bibliographystyle{IEEEtran}
\bibliography{bib}

\end{document}